\documentclass[pdflatex, sn-nature]{sn-jnl}
\usepackage{lmodern}
\usepackage[T1]{fontenc}
\usepackage[utf8]{inputenc}
\usepackage{graphicx}%
\usepackage{multirow}%
\usepackage{amsmath,amssymb,amsfonts}%
\usepackage{amsthm}%
\usepackage{mathrsfs}%
\usepackage[title]{appendix}%
\usepackage{xcolor}%
\usepackage{textcomp}%
\usepackage{manyfoot}%
\usepackage{booktabs}%
\usepackage{algorithm}%
\usepackage{algorithmicx}%
\usepackage{algpseudocode}%
\usepackage{listings}%

\begin{document}

\title{Ordering links in bicycle network plans}

\author*[1]{\fnm{Clément} \sur{Sebastiao}}\email{clse@itu.dk}

\author*[1,2]{\fnm{Michael} \sur{Szell}}\email{misz@itu.dk}

\affil[1]{\orgdiv{Networks, Data, and Society (NERDS)}, \orgname{IT University of Copenhagen}, \orgaddress{\street{Rued Langgaards Vej 7}, \city{Copenhagen}, \postcode{2300}, \country{Denmark}}}

\affil[2]{\orgname{Complexity Science Hub}, \orgaddress{\street{Metternichgasse 8}, \city{Vienna}, \postcode{1030}, \country{Austria}}}

\abstract{Many cities are announcing ambitious bicycle network plans aiming to provide safety for cyclists, with massive benefits to public health, urban livability, and the climate. A bicycle network fulfills this function if it protects from cars and if it allows reaching most of the city on direct routes and without gaps. However, having a good network plan is not enough. The temporal dimension of becoming functional throughout growth, ideally as early as possible, is crucial yet neglected in practice. To shed light on this issue, here we study Paris' 2021-2026 bicycle network plan and longitudinal data of its development stages, using the success metrics of coverage and directness. We establish that, in coverage--directness space, the city has grown its bicycle network since 2021 indistinguishably from random. Consequently, the network could have become functional earlier and is now locked into a suboptimal structure that makes it difficult to improve over random growth in the future. We systematically explore the trade-offs between twelve growth strategies, showing that strategies that either prioritize dual graph betweenness or coverage along the hierarchy of the road network are strong Pareto improvements over the real growth. Surprisingly, even just growing randomly along the road hierarchy brings substantial structural improvements. These results provide important lessons to cities that aim to become more human-centric through bicycle network development: Establish city-wide coordination and implement early major links that reach the whole city and through direct routes.}


\maketitle

\section*{Introduction}
Cycling is one of the most effective ways to boost urban livability, to provide massive public health benefits, and to substantially cut carbon emissions \cite{brandClimateChangeMitigation2021,chen_historical_2022}. Consequently, many cities have announced ambitious plans for bicycle networks in recent years. On the microscopic street design level, cyclists must be protected from the harm of cars \citep{minerCarHarmGlobal2024b}. On the macroscopic network design level, the network must be \emph{functional}: well-connected, city-covering, and providing direct routes \citep{CROW_manual}. However, it is not enough to consider only the spatial dimension. 

The temporal dimension is crucial yet neglected in practice: It can take cities decades to develop a functional bicycle network \citep{carstensenSpatiotemporalDevelopmentCopenhagens2015}, and emerging research shows that the \emph{order} in which links are built has a substantial role in arriving at a functional structure early \citep{paulsenSocietallyOptimalExpansion2023,szellGrowingUrbanBicycle2022}. Not doing so can lead to a negative spiral of perceived uselessness of the growing yet still dysfunctional network with ``bikelash'' effects \cite{wildBikelashEngagingCommunity2018} which can discourage policymakers from investing more into network development \citep{chen_historical_2022, szellGrowingUrbanBicycle2022}. Here, we fill this gap by extending and applying our recently developed framework for transport network growth \citep{sebastiaoTradeoffDirectnessCoverage2026} to the city of Paris.

Paris announced in 2021 the ``Plan Vélo 2021-2026'', a $182\,\mathrm{km}$ plan of protected bicycle lanes \cite{parisplanvelo}, expanding from $118\,\mathrm{km}$ to $300\,\mathrm{km}$, aligning with the broader urban development strategy to create a \emph{15-minute city} where essential amenities are reachable by bike or by foot in less than 15 minutes \cite{moreno15MinuteCitySolution2024}. By its 2026 deadline, the city has only implemented 43\% of its planned infrastructure, reaching $200\,\mathrm{km}$, and has done so in a seemingly uncoordinated fashion, with unequal delays between boroughs \cite{sebastiaoRightwingVotesRelate2026}. Such delays are unfortunately expected, especially in motonormative societies where cycling infrastructure is politically contested \cite{carstensenSpatiotemporalDevelopmentCopenhagens2015, walkerMotonormativityHowSocial2023, hendersonStreetFightsCopenhagen2019, wildBikelashEngagingCommunity2018}. A typical consequence is gaps in the network \cite{nateraorozcoDatadrivenStrategiesOptimal2020, vybornovaAutomatedDetectionMissing2023} that are obstructing cycling uptake \cite{schonerMissingLinkBicycle2014, ferenchakLinkLowstressBicycle2025}. A good strategy should therefore follow a city-wide perspective, to implement a functional network that is beneficial for as many citizens as possible and as early as possible throughout its construction rather than only at the end \citep{sebastiaoTradeoffDirectnessCoverage2026}. 

In this paper we study how Paris could have implemented its bicycle network plan and its options to go ahead from the current situation. We show how much more cumulative functionality the network could have gained had the city developed the network's pieces following a coordinated, city-wide approach, and how little space for improvement there is left now that its structure is locked in. In particular, our analysis shows that 1) the real network implementation was statistically indistinguishable from a random growth strategy, 2) a good greedy strategy would have been 10 standard deviations better in terms of directness and also  better in terms of coverage, 3) even the naive strategy of randomly developing along the road hierarchy could have achieved 9 standard deviations better directness. These results provide important lessons for Paris and other cities that aim to become more human-centric through bicycle network development. 

\begin{figure}
    \centering
    \includegraphics[width=0.8\linewidth]{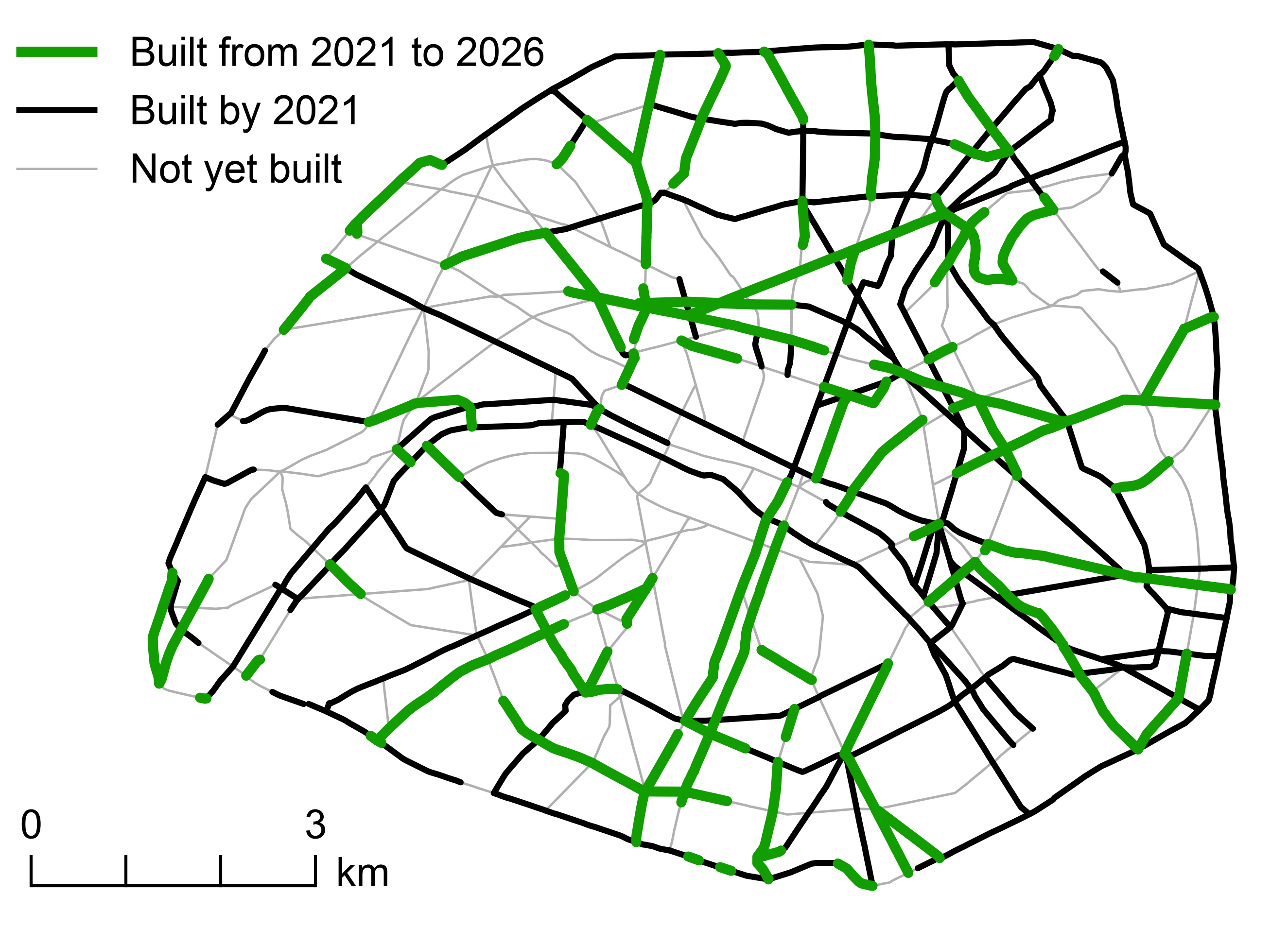}
    \caption{\textbf{Paris' ``Plan Vélo 2021-2026''.} Three exclusive progress stages of the bicycle plan are shown: 1) built by 2021 (black), 2) built from 2021 to 2026 (green), 3) not yet built but planned (grey). For an interactive version showing all intermediate growth steps, see \url{https://bikenetkit.org/sortbikenet/paris_fr/real/from_scratch}.}
    \label{fig:bikenet-plan}
\end{figure}

Paris' plan is a particularly well suited case study on how to grow bicycle networks, because longitudinal data of its growth is available between 2021 and 2026. The data (see \autoref{fig:bikenet-plan}) has been collected by a local cycling advocacy group \cite{ngobikeplan}. In our simplified network, $124\,\mathrm{km}$ had been built by 2021; from 2021 to 2026, $76\,\mathrm{km}$ more were built reaching $200\,\mathrm{km}$, with $100\,\mathrm{km}$ that are still remaining to be built. To our knowledge, there are only two other longitudinal, empirical bicycle network dataset discussed in the literature. One is also about Paris, between 2005 and 2020 \cite{moranTreatingCOVIDBike2022}. Unfortunately, the data provider since removed that dataset because of its unreliability \cite{paris_wayback}. The remaining dataset is for Copenhagen between 1913 and 2013 \cite{carstensenSpatiotemporalDevelopmentCopenhagens2015}, but that time horizon of 100 years is so long that it features strong co-evolution effects between the street network and the built environment that would excessively complicate analysis, which is not the case for the 5 years of the Paris plan. 

For details on data processing, see \nameref{sec:methods}.

We face the problem of having a bicycle network plan, which we call the \emph{final network}, for which we want to find a growth \emph{strategy} that orders the links in a way to arrive at a \emph{functional} network as early as possible during growth -- where the meaning of ``functional'' is defined by one or more network metrics. For example, if the growing network covers a large part of the city at some point, it could be called functional from a coverage perspective. This problem has been explored previously \cite{sebastiaoTradeoffDirectnessCoverage2026}, see the original reference for details on the methodological framework. This framework follows a network science based approach within the growing field of quantitative bicycle network research, as opposed to other possible approaches like GIS-based transport modelling \cite{diogopintoFrameworksPrioritizingCycling2026}.

We call a particular ordering of links an \emph{order}. If a strategy includes a random element, for example the strategy of random growth, it can produce different orders. Every such growth order starts and ends with the same network, while interesting variations happen during intermediate growth stages. Consequently, we select two performance metrics and evaluate strategies using their Area Under the Curve (AUC). The AUC aligns naturally with our objective of summarizing strategy performance across all possible levels of built length, operationalizing the idea of becoming functional as fast as possible \cite{sebastiaoTradeoffDirectnessCoverage2026}.

 \begin{figure}
    \centering
    \includegraphics[width=\linewidth]{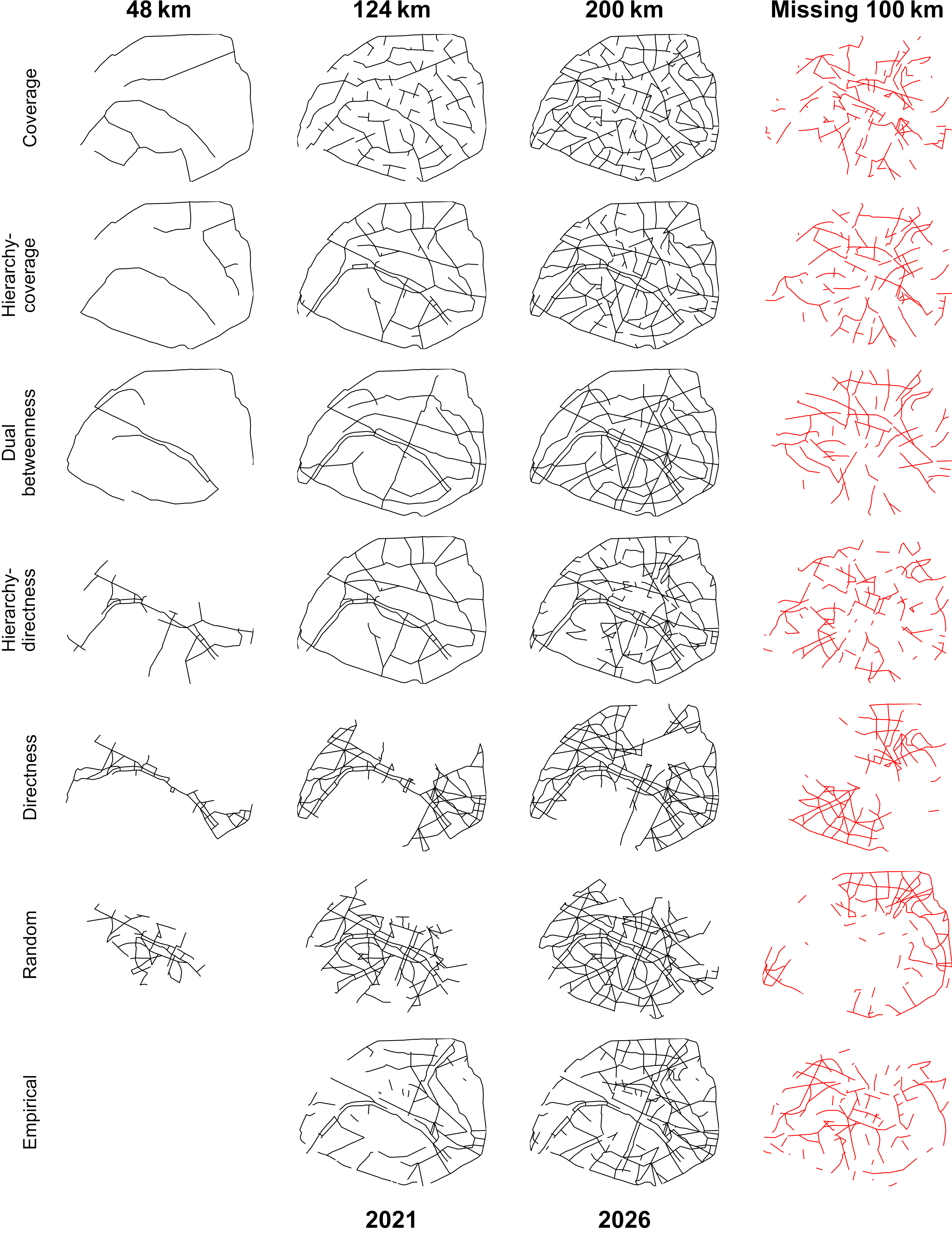}
    \caption{\textbf{Demonstration of several strategies grown from scratch.} The last row shows the empirical data at the same length in the years 2021 and 2026. The last red column shows all $100\,\mathrm{km}$ of links that are planned but missing. As we do not have empirical data before 2021, in the rest of the paper we do not consider growth from scratch, but from either the 2021 or the 2026 state. See \url{https://bikenetkit.org/sortbikenet/paris_fr/coverage/from_scratch} for an interactive version.}
    \label{fig:demo-strategies}
\end{figure}

For the two performance metrics we choose coverage and directness, as they are complementary metrics recognized as crucial by both the qualitative and quantitative literature on bicycle network design \cite{CROW_manual,schonerMissingLinkBicycle2014}. Direct comparisons between these two performance metrics are difficult because of their opposing behaviors: Directness is highly non-linear and tends to get smaller for larger networks, while coverage increases monotonically and tends to get bigger for larger networks \cite{sebastiaoTradeoffDirectnessCoverage2026}. Therefore, we are especially interested in new strategies that improve on both metrics compared to the empirical implementation, which we formalize within the framework of Pareto optimality, see \nameref{sec:methods}.

\begin{figure}
    \centering
    \includegraphics[width=\linewidth]{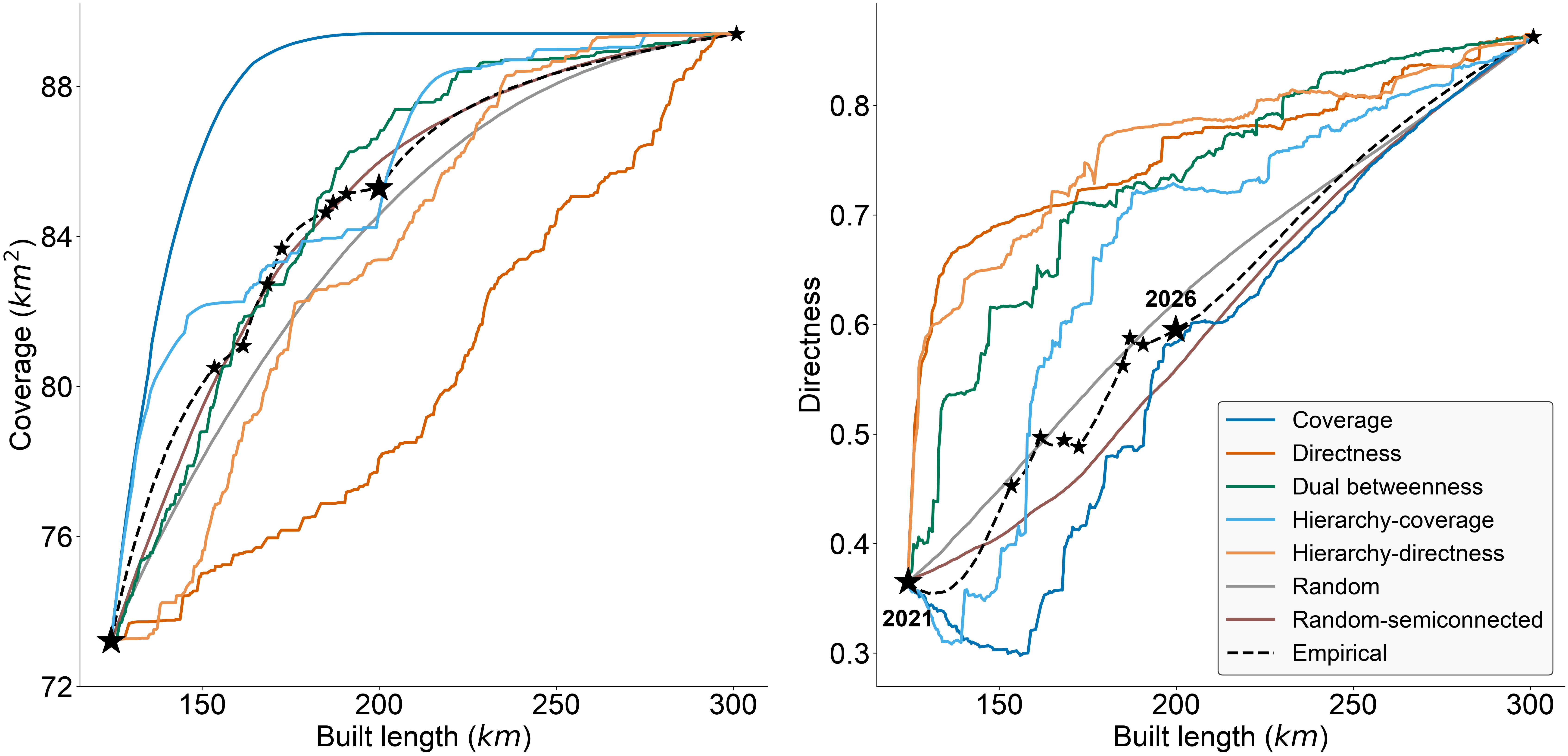}
    \caption{\textbf{Evolution of coverage and directness for selected growth orders starting from 2021.} Empirical growth (stars and dashed curves) is close to Random-semiconnected (brown) and increases its directness much slower than most greedy strategies. The areas under these curves are reported in \autoref{fig:AUC-multiple}.}
    \label{fig:lineplot-coverage-directness}
\end{figure}

\subsection*{Growth strategies}
We define twelve growth strategies and run them from the 2021 state (to know how much better Paris could have implemented the network) and from the 2026 state (to know how Paris could proceed into the future). These synthetic strategies add links only to built components, and they are all greedy, looking ahead only one step at a time as opposed to higher-order methods \cite{steinacker_higher-order_2026}. See \autoref{fig:demo-strategies} for a visual demonstration of the most distinct strategies grown from scratch, compared to the empirical states in 2021 and 2026 (bottom row). As we do not have empirical data before 2021, in the rest of the paper we do not consider growth from scratch, but from either the 2021 or the 2026 state. A detailed technical explanation of the strategies is provided in the \nameref{sec:methods}, a summary is provided below.  

\begin{figure}
    \centering
    \includegraphics[width=0.99\linewidth]{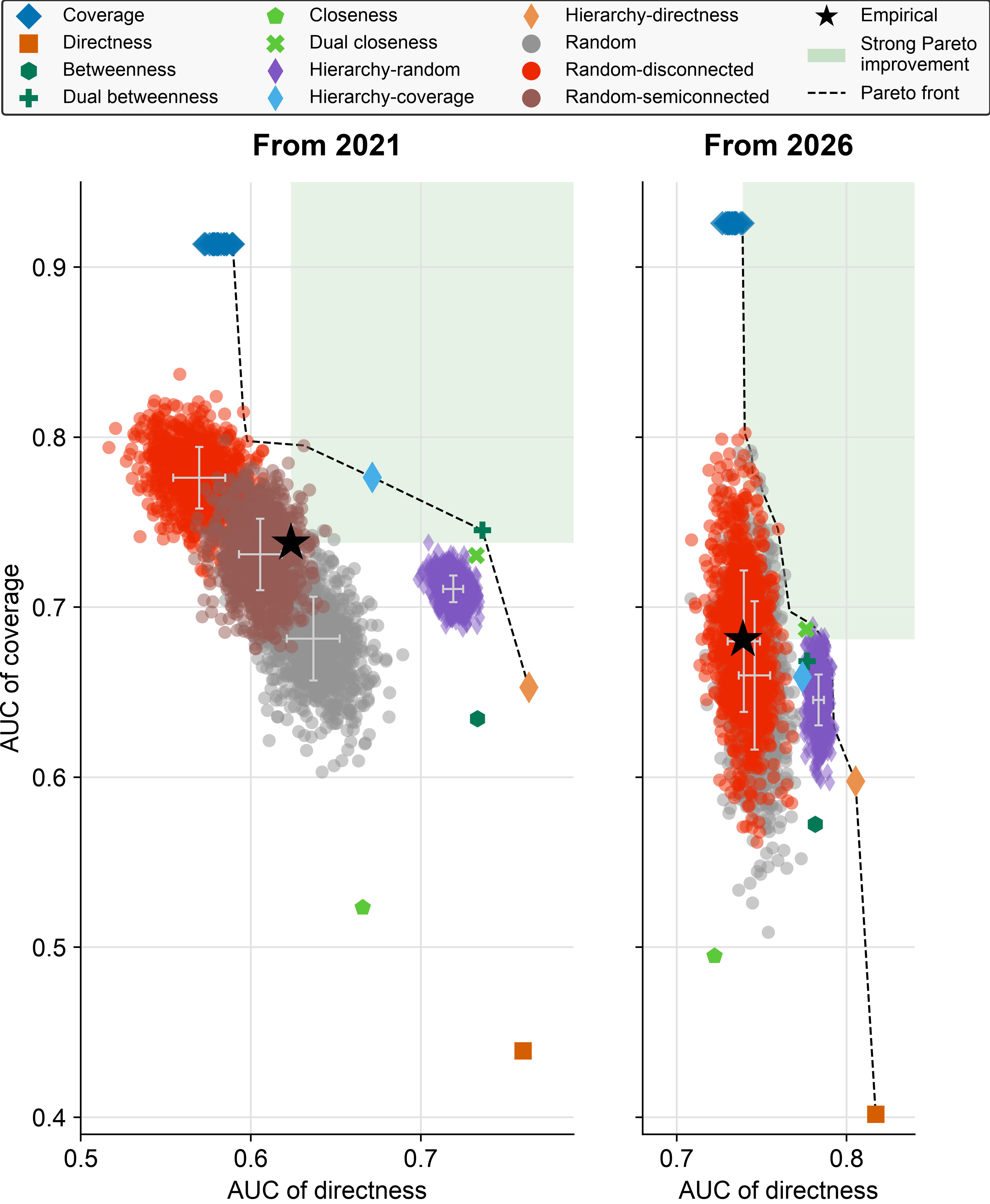}
    \caption{\textbf{Area Under the Curve (AUC) comparison of all growth orders}. The areas are taken under curves as reported in \autoref{fig:lineplot-coverage-directness}, then normalized. Growth orders start from the empirical bicycle network in 2021 (left) or in 2026 (right). Error bars for random orders and Hierarchy-random orders depicts one standard deviation from the mean over 1000 runs. Empirical growth is never on the Pareto front and is statistically indistinguishable from Random-semiconnected growth orders. When growing from 2021, two strategies are in the green area of strong Pareto improvement over the Empirical growth: 1) The Dual betweenness improves AUC of directness by over 10 standard deviations of the Random-semiconnected growth compared to empirical growth, 2) Hierarchy-coverage growth improves the AUC of coverage by over 2 and the AUC of directness by over 5 standard deviations. When growing from 2026, Random and Random-semiconnected strategies are equivalent, and greedy strategies do not provide as large improvements over random strategies any longer.}
    \label{fig:AUC-multiple}
\end{figure}

The seven simple strategies, which were already explored in Ref.~\citep{sebastiaoTradeoffDirectnessCoverage2026}, are the Coverage strategy (add the link with maximal increase in areal coverage), the Directness strategy (add the link with maximal increase in directness), the Betweenness strategy (add the link with maximal increase in betweenness centrality), the Closeness strategy (add the link with maximal increase in closeness centrality), and a random strategy (add random links) in three variations: The default Random strategy keeps the same connectedness constraint as other methods; the Random-disconnected strategy removes the connectedness constraint, allowing any link to be added at any time; the Random-semiconnected strategy allows for a new connected component in the network only as long as the next data point of the empirical network has more connected components.

On top of these, we explore the Dual betweenness and Dual closeness strategies, which work on the dual graph \cite{portaNetworkAnalysisUrban2006a}, making use of the COINS algorithm \cite{tripathyOpensourceToolExtract2021}, see \nameref{sec:methods}. The shortest path in the dual graph corresponds to the simplest path in the primal graph, i.e.~the path with the least amount of turns, which makes the Dual strategies implement long, straight roads or large ring roads first. See \url{https://bikenetkit.org/sortbikenet/paris_fr/dual_betweenness/from_scratch} for an interactive visualization.

Finally, we consider three strategies that make use of the road hierarchy, extending the Random strategy to Hierarchy-random, Coverage to Hierarchy-coverage, and Directness to Hierarchy-directness. These extensions work just as the respective original strategies, but they first consider the subnetwork of primary roads until finished, then secondary roads, then tertiary roads, then remaining roads such as residential roads. The idea behind using the hierarchy is that primary roads have already been spatially optimized by historical processes for vehicular traffic, and they are usually wider than other streets allowing easier street space reallocation \cite{basiloneRoadWidthAwareNetworkOptimisation2025,romm_measuring_2026}. These properties could make them the best candidates for adding protected bicycle infrastructure, at least from a structural, geometrical, and practical perspective.

\section*{Results}

We report the evolving coverage and directness values for selected growth strategies starting from 2021 in \autoref{fig:lineplot-coverage-directness}, left and right panels, respectively. The empirical growth (black stars connected by dashed line) shows its coverage developing in the middle between the extreme strategies that optimize for Coverage (top curve, dark blue) and for Directness (bottom curve, orange). However, its evolution of directness, \autoref{fig:lineplot-coverage-directness} right, lies for most parts below all other strategies except for Coverage (bottom curve, dark blue). 

\subsection*{Real growth performs as poorly as random growth}

We find that the empirical growth order, see black star in \autoref{fig:AUC-multiple}, is not Pareto optimal and is closer to random orders than to the Pareto front. More specifically, it is statistically indistinguishable from the Random-semiconnected orders: We establish statistical indistinguishability by switching to the AUC metric which measures the areas under these curves. For the Random-semiconnected strategy, using a Shapiro-Wilk test we cannot reject normality of the distribution of AUC of coverage (\mbox{$W=0.998$}, $p=0.25$) and of AUC of directness ($W=0.998$, $p=0.17$). Using the t-statistic we then confirm that the Empirical AUC of coverage ($t=0.32$, $p=0.75$) and directness ($t=1.44$, $p=0.15$) are indistinguishable from the Random-semiconnected strategy distribution.

Why are directness and coverage of the empirically grown network so low? The low directness is explained by the network's fragmentation, which is the most distinct characteristic of real bicycle networks \citep{nateraorozcoDatadrivenStrategiesOptimal2020,szellGrowingUrbanBicycle2022}. The coverage is higher than for the Random orders because of the higher number of connected components throughout the growth (see SI Fig. 1), as disconnected components can naturally lead to increased coverage \citep{sebastiaoTradeoffDirectnessCoverage2026}. The same phenomenon explains why the Random-semiconnected strategy is the closest to the Empirical growth, as it allows for a similar amount of connected components. We observe a continuum between the Random and the Random-disconnected orders, where the higher the number of connected components, the lower the AUC of directness and the higher the AUC of coverage.

\begin{figure}
    \centering
    \includegraphics[width=\linewidth]{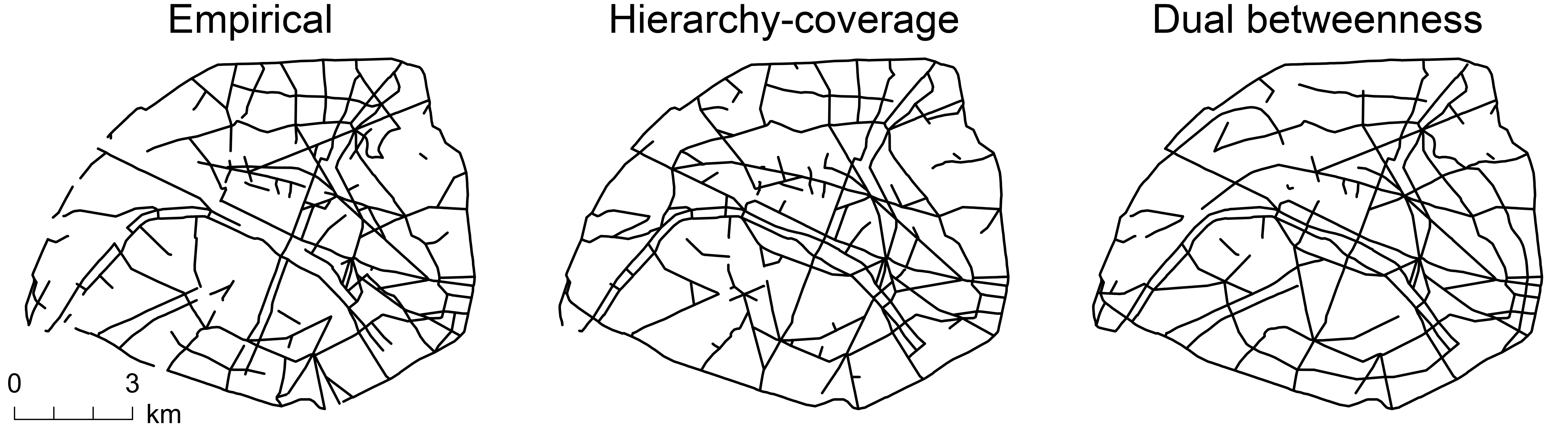}
    \caption{\textbf{Empirical growth order (left), and orders  for the synthetic growth strategies of Hierarchy-coverage (center) and Dual betweenness (right).} Shown is the state at $200\,\mathrm{km}$ in 2026. The synthetic growth orders have more continuous routes, especially along the ring road, thus less gaps. They also cover more of the city area earlier.}
    \label{fig:bikenet-growth}
\end{figure}

\subsection*{Simple strategies outperform the real growth from 2021}
We first explore the case of growth from 2021, left panel of \autoref{fig:AUC-multiple}. Four synthetic and greedy orders are on the Pareto front. This clear separation to the Empirical growth is remarkable, given these are all just naive greedy strategies that do not look ahead more than one growth step and should thus perform worse than an unrestricted human planner.

The most interesting ones are the Hierarchy-coverage and Dual betweenness strategies, as they are strong Pareto improvements over the Empirical growth. In \autoref{fig:AUC-multiple} (left) this strong improvement translates visually to the light blue diamond and the dark green cross being both positioned in the light green area to the right and upwards from the black star. Both strategies share similarities, because long roads are typically roads of both high hierarchy and high dual betweenness, and thus perform well as they focus early on the ring road and major radial streets. By forming a radio-concentric backbone of the street network, they cover most of the city and stabilize directness, as any new node will have a moderately straight path to any already added node. Dual betweenness is particularly interesting as it depends solely on the network's geometry, making it generalizable to any planar network.

\subsection*{The space for improvements has shrunk by 2026}
At the time of writing in 2026, exploring growth from 2021 is a theoretical exercise asking: How could have Paris grown its bicycle network better? However, Paris has \emph{already} grown its bicycle network a certain way until 2026, thus we are left to ask the practical question: How to grow from here? We address this question in the right panel of \autoref{fig:AUC-multiple} which depicts growth from 2026. 

Compared to the figure's left panel that shows growth from 2021, the space for improvements has shrunk considerably: The separation between random and non-random strategies is reduced. In 2026, the Random-semiconnected strategy and the Random strategy are similar. While visually the strategies are still separable, the only greedy strategy that is a strong Pareto improvement over the Empirical growth is the Dual closeness strategy, but only barely 3 standard deviations away from the Random-semiconnected average in AUC of directness. This result can be interpreted as Paris having locked in to a suboptimal network structure by 2026 that leaves much less space for improvement, at least when it comes to greedy strategies.

\subsection*{Introducing hierarchy improves the random strategy substantially}
The last result to highlight is the massive improvement from Random to Hierarchy-random, visually expressed in \autoref{fig:AUC-multiple} by the clear separation of the cloud of grey dots of Random orders and the cloud of purple diamonds of Hierarchy-random orders. For growth from 2021, \autoref{fig:AUC-multiple} (left), the separation is around 5 standard deviations in AUC of directness and 1 standard deviation in AUC of coverage. As a practical consequence, this result implies that even without \emph{any} knowledge of network measures, planners could reap massive structural improvements for directness by first implementing the bicycle network links in the highest road hierarchy, \emph{even in a random order}.

\section*{Discussion}
\label{sec:discussion}

In this article, we took on the neglected temporal dimension in bicycle network development from a network science perspective \citep{carstensenSpatiotemporalDevelopmentCopenhagens2015,paulsenSocietallyOptimalExpansion2023} (note that this temporality is less neglected from a transport modelling perspective which on the flip side tends to neglect network structure \cite{diogopintoFrameworksPrioritizingCycling2026,paulsenSocietallyOptimalExpansion2023}). We analyzed a new longitudinal dataset of an empirically growing bicycle network based on Paris' ``Plan Vélo 2021-2026''. Our analysis first established explicitly that this growth is indistinguishable from the Random-semiconnected strategy in coverage--directness space. Such a random growth hypothesis has been postulated earlier \cite{szellGrowingUrbanBicycle2022}, but evidence has so far been only implicit due to lack of longitudinal network data. This explicit evidence marks a milestone in bicycle network research.

At the same time, it is known that real growth is not completely random but biased: effects of local obstruction can play a role due to the politically contested nature of bicycle infrastructure \cite{sebastiaoRightwingVotesRelate2026}, and fragmentation can also emerge from the distributed decision processes between a diverse network of stakeholders. Achieving early functionality of a bicycle network is thus especially politically important, to counter arguments of delay like ``We already built many bike tracks but nobody is using them, so why build more?'' \cite{szellGrowingUrbanBicycle2022}. 

Protected bicycle lanes are not everything: Cities must also create the overall conditions and culture for cycling, making streets safer and more inviting to cycle on, including implementing lower speeds, traffic filters, bike-sharing programs, or reduced through-traffic to first create an environment in which cycling could flourish. For this reason, we do not give a sharp definition of the ``bicycle network'' concept, which does not have a generally agreed, context-independent definition anyway \cite{schonScopingReviewCycling2024}. We see a bicycle network as a collection of  physical or legal means that connect places safely for cyclists of all ages and demographic groups -- not just protected bicycle lanes. The microscopic implementation is deliberately left open to specific contexts and planning cultures.

By comparing an array of synthetic, greedy strategies to the empirically grown strategy, our research went beyond a mere descriptive study, exploring how sustainable transport planning could benefit massively even from naive approaches. We showed that in the cautionary example of Paris, following a connected and greedy growth strategy consistently and as early as possible would have been structurally better than what has been actually done. We identified the two best possible strategies of Dual betweenness and Hierarchy-coverage that are strong Pareto improvements of the real growth, where Dual betweenness improves around 10 standard deviations along the directness dimension. Assuming that Paris was indeed following such a Random-semiconnected strategy, this improvement implies an only 1 in 66 sixtillion chance that the city could have arrived at the same network quality. However, even the much simpler strategy of starting to \emph{randomly} build along the primary streets is surprisingly almost as effective as Dual betweenness.

We also showed the quickly diminishing opportunities in bicycle network implementation, and the importance of the early stages of growth. Starting from the partially built, $200\,\mathrm{km}$ long network in 2026 instead of the $124\,\mathrm{km}$ long network of 2021 at the start of the plan, we found the random orders in close vicinity to the greedy strategies, \autoref{fig:AUC-multiple} (right). While there are still $100\,\mathrm{km}$ to build, the suboptimal new intermediate network structure became harder to optimize, bringing the synthetic strategies to only around 3 standard deviations from Random-semiconnected orders.

Just because we did not see a better \emph{greedy} strategy does not mean that there is no better \emph{general} strategy. More advanced growth strategies based on higher order network features \cite{steinackerDemanddrivenDesignBicycle2022,paulsenSocietallyOptimalExpansion2023,steinacker_higher-order_2026}, metaheuristics \cite{martinezplumedCyclingNetworkProjects2016}, or supervised learning \cite{guo2022systematic,you2018graphrnn, kong2023autoregressive, zhou2019optimization, touwen2024learning, madeira2024generative}, could potentially create a better order and provide an idea on what the ceiling is: Is the lack of strong improvements in 2026 due to the solution space being more constrained or because greedy algorithms become worse? Nevertheless, simpler methods are better candidates to enter into praxis due to their convenience and explainability \cite{schonScopingReviewCycling2024}, especially considering how much they already outperform the current state of the art.

Our focus on two dimensions of network structure is just a first step towards practicability, as it ignores other important factors.

For example, apart from structure, traffic data and human mobility patterns, road widths \cite{basiloneRoadWidthAwareNetworkOptimisation2025,lonardiCohesiveUrbanBicycle2025}, or harder to measure human benefits like the attraction of riverside corridors \cite{kwiatkowski_why_2023} can be relevant. These are especially important in Paris, where many bicycle lanes built before 2021 run in parallel along water bodies such as the Quais de Seine, the Bassin de l'Arsenal, or the Bassin de la Villette. While these are structurally mostly redundant, they are scenic areas, increasing livability and likely to be both technically and politically easier to convert into protected bicycle infrastructure, serving more for leisure and tourism than mobility purposes -- a common strategy by municipalities \cite{kwiatkowski_why_2023}. Another structurally redundant example is Rue St.~Jaques which runs in parallel to the wide Boulevard Saint-Michel. Despite the boulevard already having a shared bus/bicycle lane, the city turned the narrow one-way street of Rue St.~Jaques after 2021 into a cycling street which provides a safer and more pleasant cycling experience.

Quantitative metrics also need to be mapped into clear goals for stakeholders to become acceptable. What is considered good or fair depends on the chosen justice theory, which translates into different metrics \cite{pereiraDistributiveJusticeEquity2017, jafinoEquitybasedTransportNetwork2021, sierts_mapping_2026}. In our case, while coverage is egalitarian as is the Gini coefficient of bicycle lanes per boroughs (see \autoref{sec:methods}), directness is closer to utilitarian.

Arguing further with the human cycling experience, the point can be made that bicycle infrastructure should be as separated from street traffic as possible, which would favor the prioritization of quiet, residential streets instead of primary roads, or place-based strategies like superblockification \cite{nieuwenhuijsenNewUrbanModels2021,buthSuperblockifyPythonPackage2024}, contrary to our results. It can also be argued that not every street needs a physically protected bicycle lane, especially for low enough levels of traffic stress. Ultimately, it does not matter how different and how much better bicycle network strategies than in reality we identify theoretically, as long as in practice political actors keep marginalizing cycling despite being one of the most effective ways to improve livability and climate \cite{brandClimateChangeMitigation2021}. 

The marginalization of cycling goes hand in hand with the lack of data collection and publication efforts, which hinders studies like ours. It was not the city of Paris that collected and published workable data on their bicycle network evolution, but a cycling advocacy group \cite{ngobikeplan}. While many cities are announcing ambitious bicycle plans, especially in the European Union as part of Sustainable Urban Mobility Plans \cite{europe_SUMP}, most cities have as public data only a schematic drawing of their plan that is not analyzable quantitatively, even less so longitudinally. Due to these reasons, our results are limited to the Paris bicycle network plan, reducing the generality of our findings. Nevertheless, given the statistically strong nature of our results, and the many similarities in bicycle network growth worldwide \cite{szellGrowingUrbanBicycle2022}, we have no reason to assume they would not reproduce similarly in other cities. Computer vision and machine learning based methods that can translate such hand-drawn maps into workable network data might enable more future studies.

Such future studies are urgently needed in the context of the triple planetary crisis, where making mobility sustainable is critical \cite{ipccAR6ClimateChange2022}. By improving not only the design but also the growth of bicycle networks towards early functionality, this research could foster a faster modal shift towards active mobility.

\section*{Methods} \label{sec:methods}

\subsection*{Data description}

Raw data of the bicycle network comes from the local cycling advocacy group ``Paris en Selle'' \cite{ngobikeplan}, who manually checked the progress of the Plan Vélo 2021-2026 on the street level at different timestamps. At the time of writing, there are 8 timestamps available: 2023-05-17, 2023-10-01, 2024-01-15, 2024-04-04, 2024-08-22, 2024-12-22, 2025-06-02, and 2026-01-28. Each timestamp come as a set of georeferenced LineStrings. We remove LineStrings in the two large public parks neighboring the city, Bois de Vincennes and Bois de Boulogne, as we focus on urban areas.

\subsection*{Data processing}

We use the road hierarchy from OpenStreetMap's (OSM) road network, extracted using \textit{OSMnx} \cite{boeingOSMnxNewMethods2017}. As OSM is highly granular, sidewalks or bicycle lanes can have their own geometry and thus their own hierarchy, we filter out non-drivable roads. We classify each road using the \emph{highway} tag. Any road that is not primary, secondary, or tertiary, we classify into a fourth level ``quaternary''. Most roads in the quaternary tier are residential, but can also be living streets, non-drivable streets, and other tags of low hierarchy that can be used by cyclists. We match the road hierarchy from OSM to the bicycle network plan data using the Hausdorff distance \cite{vieroBikeDNAToolBicycle2024}. We segmentize each LineString into shorter segments and match the target dataset to the reference dataset by minimizing Hausdorff distance between segments. 

We manually change the road hierarchy at one surgical point near the Porte de Champerret: while the bicycle lane is noted as crossing on the Avenue Stéphane Mallarmé, a tertiary road, we use the hierarchy of the nearby Boulevard Berthier, a primary road, to ensure that the ring road is only made of primary road elements.

We create a network from the resulting list of LineStrings using \textit{momepy} \cite{fleischmann_2019}. To reduce network size and increase explainability, we manually remove every dead-end and simplify the network's topology. Most of the simplification consist in merging nodes for roundabouts or plazas. For instance, we transform the large roundabout of the Place Charles de Gaulle which corresponds initially to seven nodes and links, into a single node, consistent with its structural nature of a single intersection. We merge parallel bicycle lanes along the Canal Saint-Martin due to their close proximity. 

All data processing steps are documented in the linked code repository.

\subsection*{Comparing strategies via Pareto optimality}
We compare growth orders based on Pareto optimality: For two metrics $f$ and $g$, an order $X$ is \emph{Pareto optimal} if it is not outperformed by any other order on both metrics: $\forall Z \neq X, f(Z) \leq f(X)$ or $g(Z) \leq g(X)$. An order $X$ is a \emph{strong Pareto improvement} over another order $Y$ if it performs better on \emph{both} metrics: $f(Y) < f(X)$ and $g(Y) < g(X)$. The set of Pareto optimal orders form the \emph{Pareto front}. As such, any order not on the Pareto front should not be considered viable, and we are especially interested in finding Pareto optimal strategies that are a strong Pareto improvements over the observed empirical growth (see the green areas in \autoref{fig:AUC-multiple}).

\subsection*{Connecting empirical data points}
We only have 10 data points for the empirical growth of the bicycle network between the initial bicycle network in 2021 and the final planned network. This temporal sparsity leaves open gaps. Connecting them linearly would be an inadequate oversimplification, as we know that neither coverage nor directness grow linearly throughout the growth of a transport network \cite{sebastiaoTradeoffDirectnessCoverage2026}. Therefore, we compute a more realistic interpolation by averaging 1000 disconnected random orders between data points, motivated by the random-like growth and fragmented nature of existing bicycle networks \cite{nateraorozcoDatadrivenStrategiesOptimal2020,szellGrowingUrbanBicycle2022}.
The resulting evolutions of empirical coverage and directness are depicted as dashed lines in \autoref{fig:lineplot-coverage-directness} that connect the data points (black stars).

Connecting the empirical points linearly would lead to slightly larger AUC of directness and slightly smaller AUC of coverage values that would not change any of our results qualitatively if we are growing the network from its status in 2021. If growing from 2026, the AUC becomes meaningless without interpolation as we only have two data points.

\subsection*{Metrics}
The metrics introduced in the following subsections are used as cost functions in a greedy algorithm \cite{sebastiaoTradeoffDirectnessCoverage2026}. Two of them, coverage and directness, are used as both performance metrics and as strategies, on their own or combined with the road hierarchy ranking. For the latter, the link selected is the connected link of highest available hierarchy and giving the network with the highest performance metric.

\subsubsection*{Coverage}

Coverage is the area covered by the spatially buffered links of the network. For buffering we use rounded line endings, which gives the area
\begin{align}
A(l,d) &= 2ld + 2\frac{1}{2}\pi d^2\notag\\
    &= \left(\frac{2l}{d} + \pi\right)d^2\label{eq:bufferedlink}
\end{align}
for a straight buffered link of length $l$ with buffer size $d$.

We select as a buffer size $400\,\mathrm{m}$ for all results, based on two criteria. First, it is a reasonable distance to cross without protected bicycle infrastructure, $1\,\mathrm{min}\,30\,\mathrm{s}$ at an average speed of $16\,\mathrm{km/h}$ thus 10\% of a 15-minute trip. It is also an intermediate buffer size: the fastest strategy for coverage with a buffer size of $400\,\mathrm{m}$ reaches maximum coverage value in 363 links out of 611, and 90\% of the maximum coverage in 215 links. It is also close to the median link length ($\approx415\,\mathrm{m}$). To test the robustness of our results, we tested buffer sizes ranging from $100\,\mathrm{m}$ to $800\,\mathrm{m}$ (see \nameref{sec:robustness}).

Coverage suffers from an edge effect: while some links are not necessary to reach full coverage, every link on the periphery of the network is necessary to reach full coverage, thus potentially inflating the importance of the ring road.

\subsubsection*{Directness}

We compute directness $D$ as the average of the ratio between Euclidean distance $d_E(i,j)$ and shortest network path length $d_G(i,j)$ over all node pairs $i$ and $j$:

\begin{equation}
    D =  \left< \frac{d_E(i,j)}{d_G(i,j)} \right>_{i \neq j}
\end{equation}
\noindent If node $j$ is unreachable from node $i$, $\frac{d_E(i,j)}{d_G(i,j)}=0$. The addition of this condition to the definition used previously \cite{sebastiaoTradeoffDirectnessCoverage2026} allows to take into account disconnected components.

\subsubsection*{Centralities}

Closeness centrality \cite{sabidussiCentralityIndexGraph1966} measures how close (in terms of shortest path length) a node $u$ is to any other node of the network,
\begin{equation}
    C_C(u) = \frac{N-1}{\sum\limits^{N-1}_{v \neq u}d_G(u,v)}
\end{equation}
where $N$ is the total number of nodes. Since the closeness centrality is only defined on nodes and we need to optimize by links, we compute the link closeness centrality $C^E_C(e)$ as the mean of the closeness of the nodes $(n_1, n_2)$ connected by the link $e$:

\begin{equation}
    C^E_C(e) = \frac{C_C(n_1) + C_C(n_2)}{2}
\end{equation}

Link betweenness centrality \cite{girvanCommunityStructureSocial2002} measures how frequently a link $(i,j)$ is included in the shortest paths between any pair of nodes of the network,
\begin{equation}
    C^E_B(e) = \sum\limits_{s \neq t}\frac{\sigma_{st}(e)}{\sigma_{st}} \frac{2}{N(N-1)}
\end{equation}
where $\sigma_{st}$ is the number of shortest paths between node $s$ and node $t$, and $\sigma_{st}(e)$ is the number of shortest paths between node $s$ and node $t$ that include the link $e$. The normalization factor $\frac{2}{N(N-1)}$ is the number of node pairs $\binom{N}{2}$.

\subsubsection*{Dual centralities}

In this article, we mostly use the primal graph representation of the street network in which nodes are intersections and links are road segments between them \cite{portaNetworkAnalysisUrban2006}. Another representation is the dual graph in which nodes are roads and they are connected if they intersect \cite{portaNetworkAnalysisUrban2006a}. While there are several definitions of what constitutes a road, we consider here that a road is a set of continuously aligned road segments based on Intersection Continuity Negotiation \cite{portaNetworkAnalysisUrban2006a}. Note that dual closeness centrality corresponds to ``integration'' in Space Syntax \cite{rattiSpaceSyntaxInconsistencies2004}.

To create the dual graph, we use the COINS algorithm \cite{tripathyOpensourceToolExtract2021} implemented in \textit{momepy} \cite{fleischmann_2019}. To define the dual graph, the COINS algorithm joins at each intersection road segments that have the largest interior angle between them. Note that in \textit{momepy}, the dual graph created by this method is called ``continuity graph'' to avoid confusion with another possible definition of dual graph.

\subsection*{Robustness analysis}
\label{sec:robustness}
We assessed the robustness of our results in the choice of parameters and in the definition of our metrics.

First, we evaluated the effect of the chosen buffer size. We explored buffer size values between $100\,\mathrm{m}$ and $800\,\mathrm{m}$, as our chosen buffer size is $400\,\mathrm{m}$ (see SI Figs. 7 to 35). When the buffer size is too small, the variations in AUC of coverage are small (only from $0.48$ to $0.62$), as the growth in coverage becomes almost linear for all growth strategies. When the buffer size is too large, most of the coverage is already reached with the bicycle network from 2021 ($97\%$ for a buffer size of $800\,\mathrm{m}$), reducing the relevance of the metric.

Coverage is one way to measure the spatial coverage. We can also measure the amount of infrastructure in each of the boroughs of Paris as a metric capturing equity. We define a Gini coefficient $G$ by measuring the number of kilometers of bicycle infrastructure in each borough, normalized by their area. We find that there is a small central bias, with more central boroughs having on average a higher share of bicycle infrastructure (see SI Fig. 6), but that the amount of infrastructure is overall equally distributed ($G=0.11$). There is a strong correlation between the AUC of Coverage and the AUC of Gini (Pearson $\rho = 0.93$, $p=1.4*10^{-6}$), making the metric redundant (see SI Figs. 4 and 5).

As we care about the performance of the bicycle infrastructure over time, we use the AUC. To account for the fact that infrastructure built earlier exist for longer and thus has higher cumulative benefits, we can add exponential discounting to the AUC: 

\begin{equation}
    f(x) = x\text{e}^{-x\text{ln}(\Delta)}
\end{equation}

We find that for $\Delta = 10$ (meaning that the first link is $10$ times more valuable than the last one), while the absolute values of AUC of coverage and directness diminish, the relative position of each order stays stable (see SI Figs.~2 and 3).

\section*{Data and code availability}

We make publicly available all code and processed data in the following repository: \url{https://github.com/csebastiao/paris_orderbike}. Road network data is available on OpenStreetMap. Bicycle network plan progress raw data is available on the ``Observatoire du Plan Vélo'' \cite{ngobikeplan}. Each timestamp is accessible at the following URLs: \url{https://observatoire.parisenselle.fr/map-YYYY-MM-DD.json} where YYYY-MM-DD is replaced by the timestamps. A Python library based on the code is being developed at: \url{https://github.com/BikeNetKit/SortBikeNet}.

\section*{Acknowledgments}

We thank all data providers, especially Paris en Selle who also provided useful correspondence, as well as Anastassia Vybornova and Martin Fleischmann for technical support in dual graph computations. C.S.~and M.S.~acknowledge support by the European Union through the Horizon Europe grant JUST STREETS (Grant agreement ID: 101104240).

\section*{Author contributions}
\textbf{Clément Sebastiao:} Conceptualization, Formal analysis, Investigation, Methodology, Software, Visualization, Writing – original draft, Writing – review and editing. \textbf{Michael Szell:} Conceptualization, Funding acquisition, Methodology, Project administration, Supervision, Visualization, Writing – original draft, Writing – review and editing.

\bibliography{references}

\end{document}